# Field effect and local gating in nitrogen-terminated nanopores (NtNP) and nanogaps (NtNG) in graphene

*Ivana Djurišić[1], Miloš S. Dražić[1], Aleksandar Ž. Tomović[1], Marko Spasenović[2], Željko Šljivančanin[3], Vladimir P. Jovanović[4] and Radomir Zikic[1,5]\**

[1]Institute of Physics, University of Belgrade, Pregrevica 118, 11000 Belgrade, Serbia.

[2]Center of Microelectronic Technologies, Institute of Chemistry, Technology and Metallurgy, University of Belgrade, Njegoševa 12, 11000 Belgrade, Serbia.

[3]Vinča Institute of Nuclear Sciences, University of Belgrade, Mike Petrovića Alasa 12-14, 11000 Belgrade, Serbia.

[4]Institute for Multidisciplinary Research, University of Belgrade, Kneza Višeslava 1, 11000 Belgrade, Serbia.

[5]NanoCentre Serbia, Nemanjina 22-26, 11000 Belgrade, Serbia.

KEYWORDS graphene, nanogap, nanopore, DFT+NEGF, field effect, termination

ABSTRACT: Single-molecule biosensing, with a promise of being applied in protein and DNA sequencing, could be achieved using tunneling current approach. Electrode-molecule-electrode tunneling current critically depends on whether molecular levels contribute to electronic transport or not. Here we found employing DFT and Non-Equilibrium Green's Function formalism that energies of benzene molecular levels placed between graphene electrodes are strongly influenced by electrode termination. Termination-dependent dipoles formed at the electrode ends induce in-gap field effect that is responsible for shifting of molecular levels. We show that the HOMO is closest to Fermi energy for nitrogen-terminated nanogaps (NtNGs) and nanopores (NtNPs), promoting them as strong candidates for single-molecule sensing applications.

Single-molecule probing by tunneling currents is often a method of choice for biomolecule sensing applications, most prominent being next-generation real-time single-molecule DNA and protein sequencing [1,2]. Tunneling current is extremely sensitive to position of a molecule placed in the nanogap between two nanoelectrodes and to its composition. Discrimination between molecules is based on variations of the tunneling current induced by interaction of a molecule traversing the nanogap with electrodes. The molecule-electrodes interaction will strongly depend on the choice of the electrodes [3, 4], as it may improve transport properties through molecules placed between them by inducing resonant transport – highest occupied/lowest unoccupied (HOMO/LUMO) molecular levels contribute to the current. Tunneling current may vary several orders of magnitude depending on whether molecular levels participate in the transport or not. Resonant transport is commonly accomplished by passivation of electrodes either through weak interaction between the molecule and electrodes (hydrogen bonding or Van der Waals) [5,3,6] and/or by electrode interface dipoles induced electrostatic field [7,8].

Particularly important material for single-molecule probing applications is graphene [1,2,9,10,11]: it is atomically thin which significantly increases resolution of devices; it can be utilized for transversal electrodes; it is impenetrable to ions and thus can be used as a membrane with nanopore; it is easily produced and integrated into devices. Most commonly, graphene edges are passivized (functionalized, terminated) with hydrogen atoms, however different terminations of graphene both simple (single atom passivation, nitrogen or hydrogen) [3,12,13] and more complex, with different recognition molecules [4,6,14] were used in order to achieve resonant transport.

In our previous study we have found strong in-gap field effect generated by dipoles formed at the ends of nitrogen-terminated carbon-nanotube electrodes in a DNA sequencing device [8]. Here, in this study we investigate the field effect in graphene nanogaps and nanopores. The influence of $X$-terminated ($X$ = N, H, F, S, Cl) semi-infinite graphene nanogap on the position of highest occupied molecular orbital HOMO of the benzene molecule in respect to Fermi energy $E_F$ was first explored. We showed that the position of HOMO level of the molecule in respect to $E_F$ is mainly determined by in-gap electrostatic potential which is formed as a consequence of termination-induced dipoles. It is found that N termination is most favorable in the sense that it drives molecular HOMO level closest to $E_F$ compared to other investigated terminations. We also showed that N-terminated grapehene nanopore should increase the energy levels of a molecule placed in the pore by the same mechanism.

Calculations for empty nanogaps with different terminations and benzene molecules in such gaps were performed using density functional theory (DFT) coupled with non-equilibrium Green's functions (NEGF) within TranSIESTA package [15]. Positions of $X$ ($X$=N, H, F, S, Cl) atoms that terminate graphene nanogaps are shown in Figure 1a. The distance between two graphene sheets, excluding termination is fixed at 11.7 Å (arrow in Figure 1a). Geometry of benzene molecule perpendicular (⊥) to the plane of the electrodes is given in Figure 1b, while parallel (∥) orientation is given in Figure S1 in Supporting Information (SI). The central region, also called the scattering region or the extended molecule is marked with dashed red rectangle (Figure 1b). Bulk electrodes are represented by six rows of carbon atoms. Unit cell for DFT calculations of N-terminated graphene nanopore within SIESTA package [16] is presented in Figure 1c.

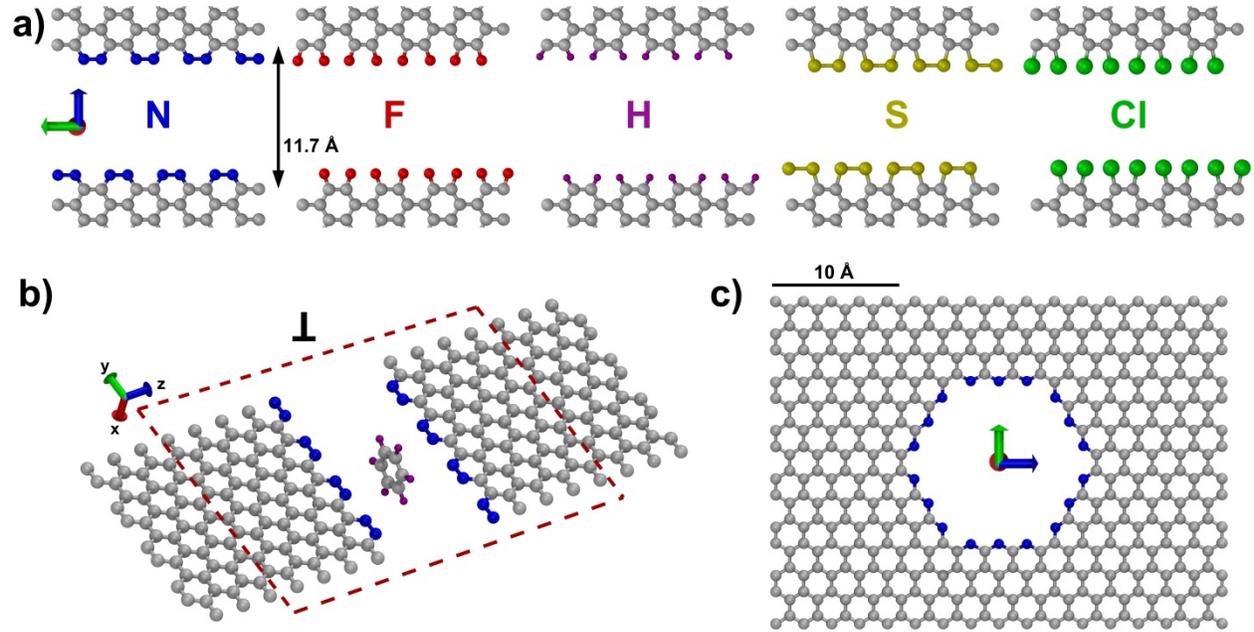

**Figure 1.** a) Geometry of *X*-terminated graphene gaps: *X*=N, H, F, S, Cl. Arrow indicates the fixed distance between C atoms adjacent to termination atoms. b) N-terminated gap with benzene molecule in perpendicular (⊥) orientation. Dashed red line marks the central region (the scattering region or the extended molecule). c) Geometry of N-terminated graphene nanopore (NtNP). Red, green and blue arrows indicate x, y and z directions, respectively.

For both SIESTA and TranSIESTA the basis set was double zeta polarized for all atoms, the exchange-correlation functional was approximated with the Perdew-Burke-Ernzerhof (PBE) functional [17], core electrons were described with Troullier-Martins norm-conserving pseudopotentials [18] and the mesh cutoff value was 170 Ry for the real-space grid. Charge excess per atom was obtained from Hirshfeld population analysis. Geometries of benzene molecule and *X*-termination atoms at the edges of graphene electrodes and graphene nanopores, were relaxed using SIESTA. For electrode and scattering region calculations in TranSIESTA we used 1×9×64 and 1×9×1 k-points, respectively. Electrostatic potential energy profiles of N and H-terminated graphen nanopores were also calculated using GPAW code [19] based on the real space projector augmented waves method [20,21]. Grid spacing for the wavefunction representation was 0.15 Å, exchange-correlation functional was PBE within generalized gradient approximation and only the Γ-point Brillouin-zone sampling was used. The electronic states were occupied according to a room-temperature Fermi-Dirac distribution.

The energy difference $E_{HOMO}^{\perp}(X)$, obtained from TranSIESTA, between molecular HOMO energy of benzene placed between *X*-terminated electrodes and Fermi energy $E_F$ depends on termination *X* (Figure 2a and Figure S2a in SI). No matter the orientation of benzene molecule (perpendicular or parallel) within the nanogap, HOMO energy is closest to $E_F$ for N termination; then F, H, S and Cl terminations follow, respectively.

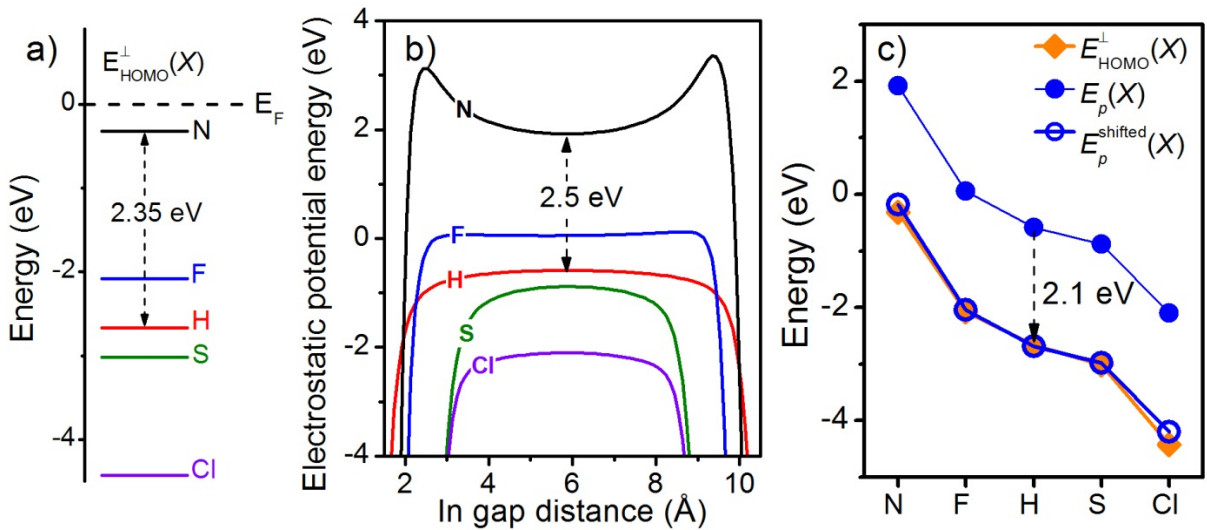

**Figure 2.** a) Energy of HOMO level with respect to Fermi energy, $E_{HOMO}^{\perp}(X)$, for perpendicularly oriented benzene (see Figure 1b) in the nanogaps with different terminations (N-

black, F-blue, H-red, S-olive and Cl-violet). b) In-gap electrostatic potential energy $E_p(X)$ along midline in z direction in electrode plane for *X*-terminated gaps obtained from TranSIESTA. c) Comparison of $E_{HOMO}^{\perp}(X)$ (solid orange diamonds) and $E_p(X)$ (solid blue circles) in the center of the gap for different terminations. Empty blue circles correspond to $E_p(X)$ values in the center of the gap shifted for -2.1 eV.

Next we calculate, from TranSIESTA, electrostatic potential energy $E_p$ for *X*-terminated empty graphene gaps, whose profiles (along midline in z direction in electrode plane) are given in Figure 2b. The values of $E_p$ in the center of the gap shifted for -2.1 eV vary in the same way as $E_{HOMO}^{\perp}$ with termination *X* (Figure 2c). Thus, the position of the molecular HOMO energy is primarily determined by the electrostatic potential of empty gap: we will define this as *field effect* similar to the one in solid-state electronics with the difference that here no gate electrode is needed. From the application viewpoint the most favorable is to use N-terminated nanogaps (NtNGs), as molecular HOMO is brought closest to Fermi energy for nitrogen termination. This termination is expected to provide the strongest current signals at lowest possible voltages if no strong interaction (hybridization) between molecule and the electrodes is present [8].

The largest value of the difference between HOMO and Fermi energies of benzene is for Cl termination (~4.1 eV), which is less than molecular energy gap (~5.2 eV). Thus, in situations where HOMO energy variation with the type of termination is smaller than molecular gap, one may expect this rule of thumb: $E_{HOMO}$ will be closest to $E_F$ for N- and then follow F-, H-, S- and Cl-terminations, respectively. However, nothing can be said *a priori* about the exact value of $E_{HOMO}$ with respect to $E_F$, as it may vary from molecule to molecule.

Similarly, in the case of parallel orientation of benzene, the position of HOMO energy with respect to $E_F$, $\text{E}_{\text{HOMO}}^{\parallel}$, is mainly determined by the values of electrostatic potential energy $E_p$ in the center of the empty gap (Figure S2b in SI). However, there is notable upshift of values of $\text{E}_{\text{HOMO}}^{\parallel}(X)$ when compared to $\text{E}_{\text{HOMO}}^{\perp}(X)$ for all terminations except H-termination (Figure S2a in SI), where $\text{E}_{\text{HOMO}}^{\parallel}(\text{H}) = \text{E}_{\text{HOMO}}^{\perp}(\text{H})$. This is expected from the point of view of coupling between benzene and termination atoms: in both cases of orientation there is negligible coupling. If we take charging of the molecule (Figure S3 in SI) and charge redistribution between termination atoms and benzene (Figure S4 in SI) to be the measure of the coupling, then coupling is much stronger in parallel than in perpendicular orientation for all terminations except H termination. Combined effect of local change of electrostatic potential due to charge redistribution, as well as charging of the molecule (negative charge rises energy) leads to upshift of HOMO level in the case of parallel orientation (Figure S2 in SI).

We will now show, using a simple model, that termination-dependent in-gap electrostatic potential energy, i.e. the HOMO shift, originates from dipoles formed at the interfaces of graphene sheets [8], as depicted in Figure 3. From TranSIESTA Hirshfeld population analysis, dipoles are seen as negative charge accumulated on N-termination row, while positive charge is located on the adjacent row of C atoms (Figure 3a), for the exemplar case of empty N-terminated graphene nanogap. The presence of benzene in the gap does not significantly alter charge distribution and dipole orientations in graphene (Figure S5 in SI).

Simple model for calculation of electrostatic potential energy consists of four homogeneously linearly charged lines (Figure 3b). Each line represents one row of atoms (blue – termination, green – adjacent C atoms) of graphene sheet. Position and length of a line correspond to position of row of atoms and the distance between the first and the last atom of a row, respectively.

Linear charge densities λ(C) and λ(X) were derived as total charge excess in one graphene row calculated from Hirshfeld analysis divided by the distance between the first and the last atom of that row. Calculated in-gap electrostatic potential energy along midline in z direction in the plane containing linear charges for different terminations is given in Figure 3c. Our model shows good qualitative agreement with results obtained from TranSIESTA calculation for empty gap (Figure 2b), except for F termination. The reason for this discrepancy is that for F termination there is significant contribution to $E_p$ from second row of C atoms counting from termination (Figure S5 in SI). If we take the model with six lines of charges (Figure S6 in SI) better qualitative agreement is obtained. The simple model implies that the electrostatic potential energy calculated from TranSIESTA emanates from dipoles, i.e. that the field effect comes from local gating induced by properly chosen termination of graphene sheet.

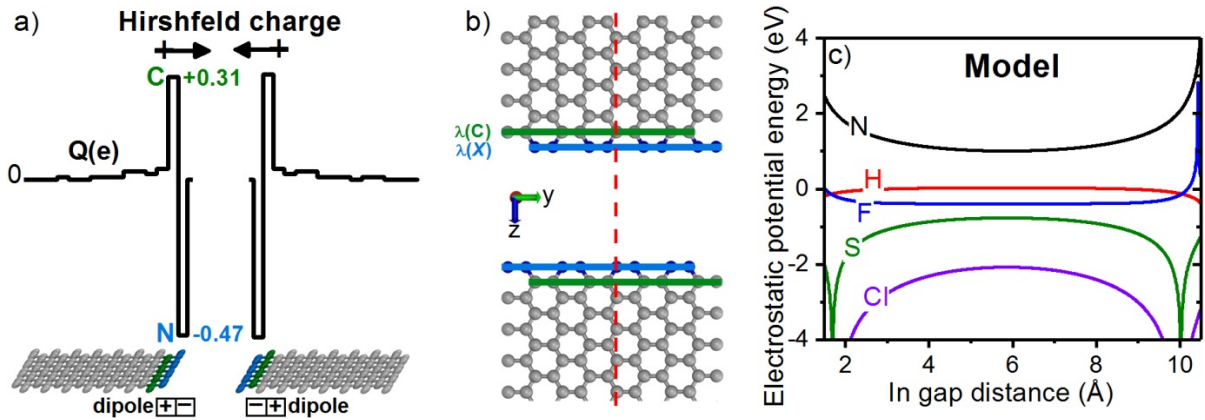

**Figure 3.** a) Hirshfeld charge excess $Q$ for each row of atoms of gap-forming N-terminated graphene sheets at zero bias (solid black line). The dipole arises in the N (blue) and the adjacent C (green) atom rows. Arrows indicate the orientation of dipoles. b) A model of four homogenously linearly charged lines represents corresponding rows of atoms (blue – termination, green – adjacent C atoms) belonging to graphene sheets. Total charge residing on the line is equal to total charge excess of corresponding row. Position and length of a line

correspond to position of and distance between the first and the last atom of a row, respectively. c) Electrostatic potential energy ($E_p$) calculated from the four-line model for different terminations along dashed red line in b), which lies in y-z plane.

The model may indicate how the variation of charge within dipoles can influence the electrostatic potential energy, i.e. HOMO level shift. For a "symmetric" dipole, $\lambda(X) = -\lambda(C)$, the field effect is stronger for a larger dipole moment (Figure S7a in SI). Negative $\lambda(X)$ (solid lines in Figure S7a in SI) produces positive electrostatic energy and a molecule placed in the nanogap would experience a positive shift of energy levels (Figure S7c in SI). The opposite is true for $\lambda(X) > 0$ (dotted lines in Figure S7a in SI). The "asymmetric" dipole case $\lambda(X) \neq -\lambda(C)$ is more realistic description of *X*-terminated graphene. Electrostatic potential energy will be positive (negative) whenever the total charge, $\lambda(X) + \lambda(C)$, of the dipole is negative (positive), irrespective of orientation of interface dipole (Figure S7b,c in SI). The larger the total charge of the dipole (asymmetry) the larger the field effect (Figure S7c in SI). The gap size and the distance between dipole-forming lines can also influence the behavior of the energy, but to lesser extent.

We demonstrated that the field-effect and local gating are the most favorable for N-terminated graphene nanogaps (NtNG). Now we will show that field effect is also present in N and H-terminated graphene nanopores (NtNP in Figure 1c and HtNP in Figure S8 in SI). Electrostatic potential energy $E_p$ obtained from SIESTA calculation for NtNP is presented in Figure 4a: in the nanopore $E_p$ is positive implying the shift of molecular HOMO level towards Fermi energy if a molecule is placed in the NtNP. The results for $E_p$ of NtNP is in good agreement with the one obtained using GPAW code (Figure 4b). Additionally, by analogy with results of calculations for terminated nanogaps, the values of $E_p$ in the HtNP are lower than in NtNP as expected (Figure 4b). As in NtNG case, the charge distribution obtained from Hirdhfeld analysis in SIESTA at the N termination and next-neighbor C atoms of the NtNP (Figure 4c) is responsible for the hat-shaped potential energy: N atoms bear negative, while next-neighbor C atoms carry positive charge, forming a ring of radially oriented dipoles. Induced electron density for NtNP obtained using GPAW (Figure S9 in SI) supports the result of Hirshfeld analysis. Hat-shaped potential can be obtained from a model of two linearly charged rings (Figure 4d), which is in good qualitative agreement with both SIESTA and GPAW calculations.

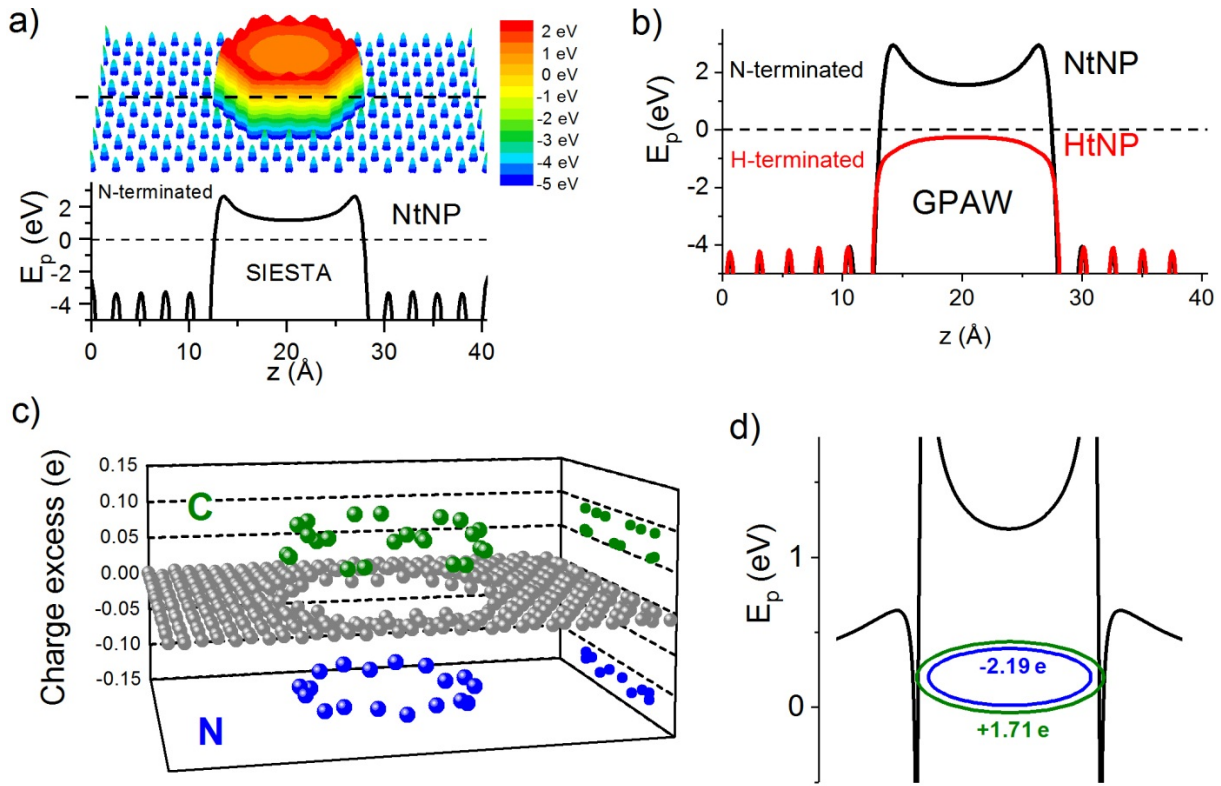

**Figure 4.** a) 3D profile (top panel) of electrostatic potential energy $E_p$ in y-z plane of NtNP (Figure 1c) and its profile (bottom panel) along the midline (dashed line in top panel) in z

direction in the same plane obtained from SIESTA. b) Electrostatic potential energy along the same line as in a) for NtNP and HtNP obtained using GPAW. c) Hirshfeld charge excess $q$ for each atom in NtNP from Figure 1c: N atoms (blue balls), their next-neighboring C atoms (green balls) and the rest of C atoms (grey balls). Projections of charge to $q$-y plane are given for N and neighboring C atoms. d) Electrostatic potential energy profile obtained from two linearly charged rings along the axis defined by their diameter. Total charge residing on rings is equal to total charge excess of N and C atoms from c).

This local-gating field effect in NtNGs and NtNPs could improve operation of single-molecule biosensing devices based on transversal current by lowering their operating voltages as molecular HOMO should be drawn closer to $E_F$. If one applies NtNGs, transversal tunneling current measured is on the order of nA or less and expected ratio between resonant (molecular HOMO contributes to electric current) and non-resonant (no molecular levels contribute to electric current) current would be much larger than 1. In applications with NtNPs one measures in-plane current that is on the order of μA [11,13]. Inclusion of molecular HOMO in transport only slightly affects in-plane current (the ratio between resonant and non-resonant current is expected to be close to but different than 1) however such small variation should be measurable due to its magnitude [11].

We have shown that N termination of graphene nanogaps tends to shift HOMO level of the molecule closest to $E_F$ when compared to F, H, S and Cl terminations. This behavior is a consequence of in-gap field effect which arises from interface dipoles formed by graphene sheet termination. In-gap field effect due to local gating by termination is also demonstrated for N-terminated graphene nanopores. From practical standpoint this means that is more likely that resonant transport will be achieved for lower values of applied voltage. We therefore suggest that for molecular biosensing based on electric transport measurements one should use N-terminated graphene nanogaps and nanopores as they could be achieved through room-temperature edge functionalization in mild $NH_3$ plasma [ 22].

ACKNOWLEDGMENT

This work was supported by the Serbian Ministry of Education, Science and Technological Development through projects 171033 and 41028. We gratefully acknowledge financial support from the Swiss National Science Foundation (SCOPES project No.52406) and the FP7-NMP, project acronym *nanoDNAsequencing*, GA214840.

ABBREVIATIONS

DFT, density functional theory; NEGF, Non-Equilibrium Green's Function; NtNP, nitrogen-terminated nanopores; NtNG, nitrogen-terminated nanogaps; HOMO, highest occupied molecular orbital; LUMO, lowest unoccupied molecular orbital.

ASSOCIATED CONTENT
**Supporting Information**. The following files are available free of charge.

Supporting Information (PDF)

**Corresponding Author**

Radomir Zikic radomir.zikic@ncs.rs

# Supporting Information

Field effect and local gating in nitrogen-terminated nanopores (NtNP) and nanogaps (NtNG) in graphene

*Ivana Djurišić[1], Miloš S. Dražić[1], Aleksandar Ž. Tomović[1], Marko Spasenović[2], Željko Šljivančanin[3], Vladimir P. Jovanović[4] and Radomir Zikic[1,5]\**

[1]Institute of Physics, University of Belgrade, Pregrevica 118, 11000 Belgrade, Serbia.

[2]Center of Microelectronic Technologies, Institute of Chemistry, Technology and Metallurgy, University of Belgrade, Njegoševa 12, 11000 Belgrade, Serbia.

[3]Vinča Institute of Nuclear Sciences, University of Belgrade, Mike Petrovića Alasa 12-14, 11000 Belgrade, Serbia.

[4]Institute for Multidisciplinary Research, University of Belgrade, Kneza Višeslava 1, 11000 Belgrade, Serbia.

[5]NanoCentre Serbia, Nemanjina 22-26, 11000 Belgrade, Serbia.

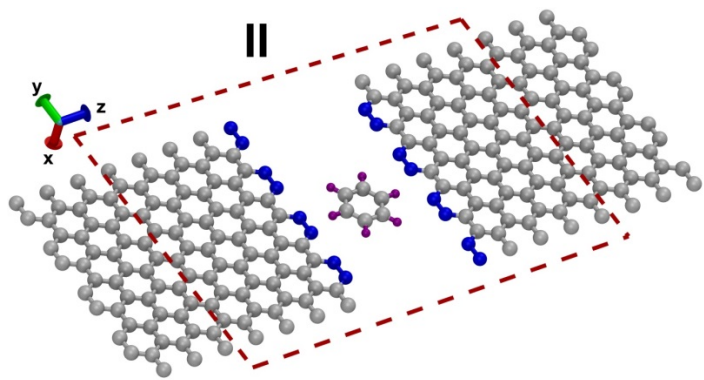

**Figure S1**. N-terminated gap with benzene molecule in parallel (∥) orientation. Dashed red line marks the central region.

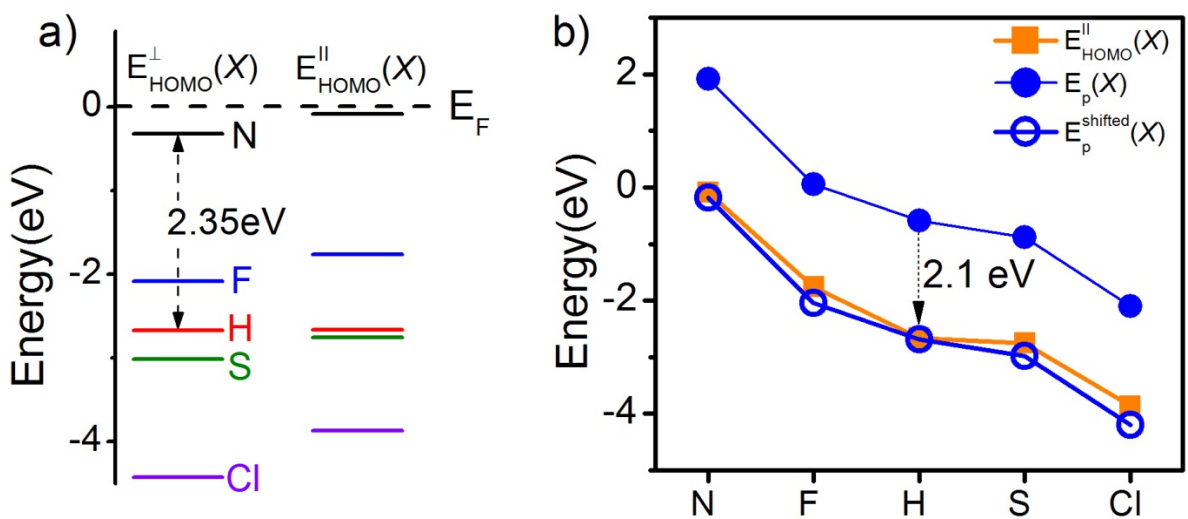

**Figure S2**. a) Position of benzene HOMO level for perpendicular (⊥) and parallel (∥) orientation and different terminations of graphene (N-black, F-blue, H-red, S-olive and Cl-violet). b) Comparison of $E_{HOMO}^{\parallel}(X)$ (solid orange squares) and $E_p(X)$ (solid blue circles) in the center of the gap for different terminations $X$. Empty blue circles correspond to $E_p(X)$ values in the center of the gap shifted for -2.1 eV.

Positions of benzene HOMO level for perpendicular (⊥) and parallel (∥) orientations and different terminations are given in Figure S2a. For all terminations HOMO energies are closer to Fermi energy $E_F$ in the case of parallel orientation, except H for which $E_{HOMO}^{\parallel}(H)=E_{HOMO}^{\perp}(H)$. This upshift of HOMO level is a result of coupling between termination atoms and benzene molecule. Coupling leads to charge redistribution between termination atoms and benzene, which results in the charging of the molecule (Figure S3).

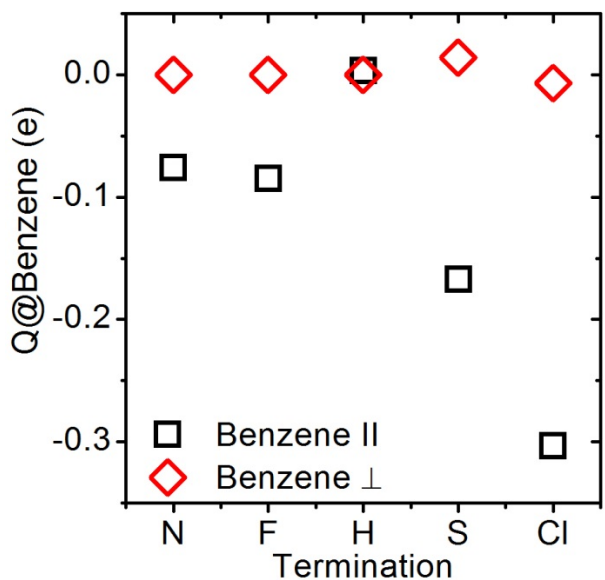

**Figure S3**. Hirshfeld charge excess residing on the benzene molecule for perpendicular (open red diamonds) and parallel (empty black squares) orientation and different terminations.

Charge residing on the benzene molecule for ∥ and ⊥ orientation is given in Figure S3. For ⊥ orientation benzene molecule is electroneutral for N, F and H, while for S and Cl-termination there is some charge residing at the molecule. This is expected as C-S and C-Cl bond lengths are larger than for other terminations (Figure 1a). In the case of ∥ orientation, charging, i.e. coupling exists for all terminations expect H termination (Figure S3).

Hirshfeld charge excess distribution for the case of N-terminated gap is given in Figure S4a. Charge distribution is perturbed at the nearest termination atoms when benzene molecule is introduced in the gap (Figure S4b,c). In perpendicular orientation, as consequence of negligible coupling, charge residing at benzene molecule is zero while nearest termination atoms are only slightly perturbed. In the case of parallel orientation (Figure S4c) the situation is drastically different: most of the charge redistribution is between benzene molecule and nearest atoms of termination.

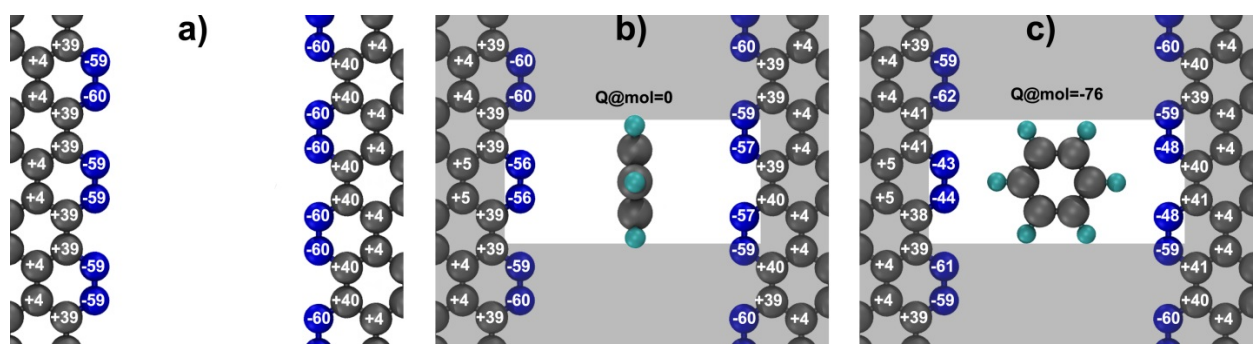

**Figure S4**. Hirshfeld charge excess per atom given in $10^{-3}$ e for graphene gaps: a) empty, b) with benzene molecule in perpendicular and c) parallel orientation.

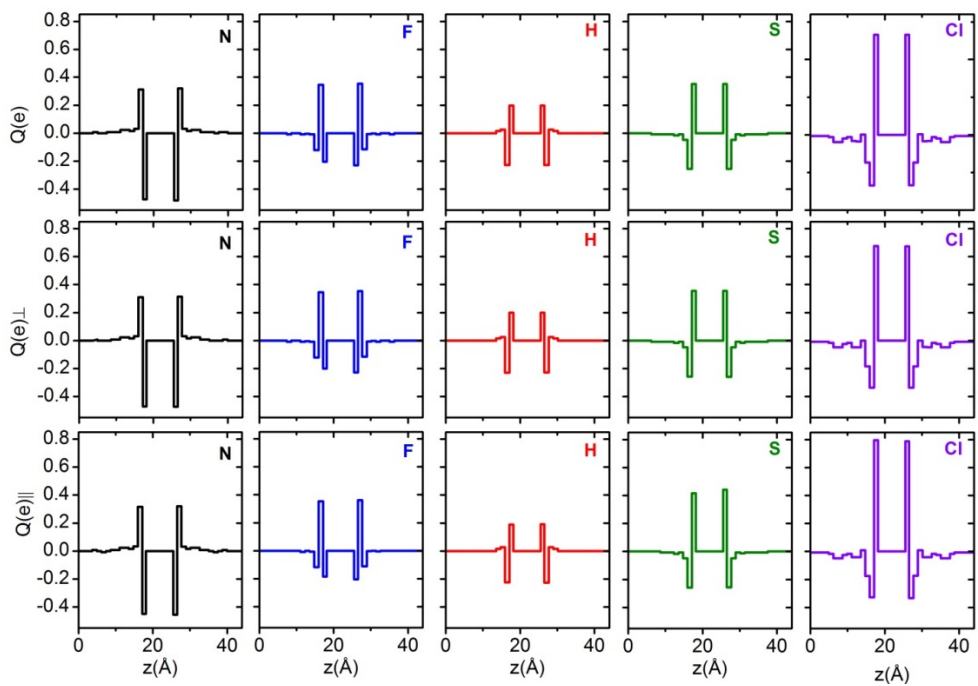

**Figure S5**. Sum of atomic charge excesses residing at each row (eight atoms along y axes) of XtNG, obtained from Hirshfeld population analysis for *X*-terminated empty gaps (top panel) and gaps containing benzene molecule in perpendicular (middle panel) and parallel (bottom panel) orientation.

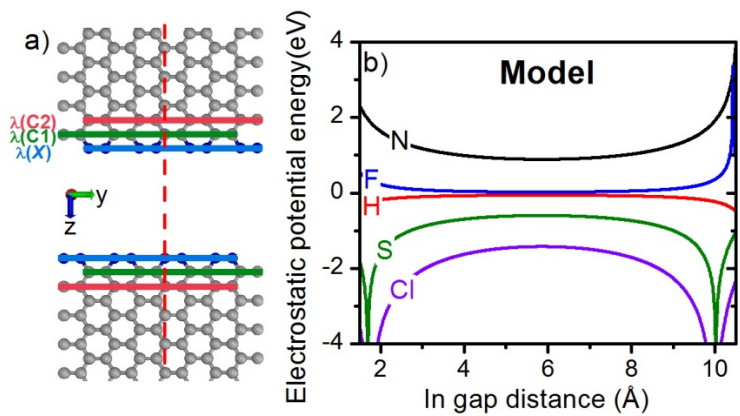

**Figure S6**. a) A model of six homogenously linearly charged lines represent corresponding rows of atoms (blue – termination, green – first and red – second row of C atoms) belonging to graphene sheets. Total charge residing on the line is equal to total charge of corresponding row.

Position and length of a line correspond to position and distance between first and last atom of a row, respectively. c) Electrostatic potential energy ($E_p$) calculated from the six-line model for different terminations along dashed red line in b), which lies in x-y plane.

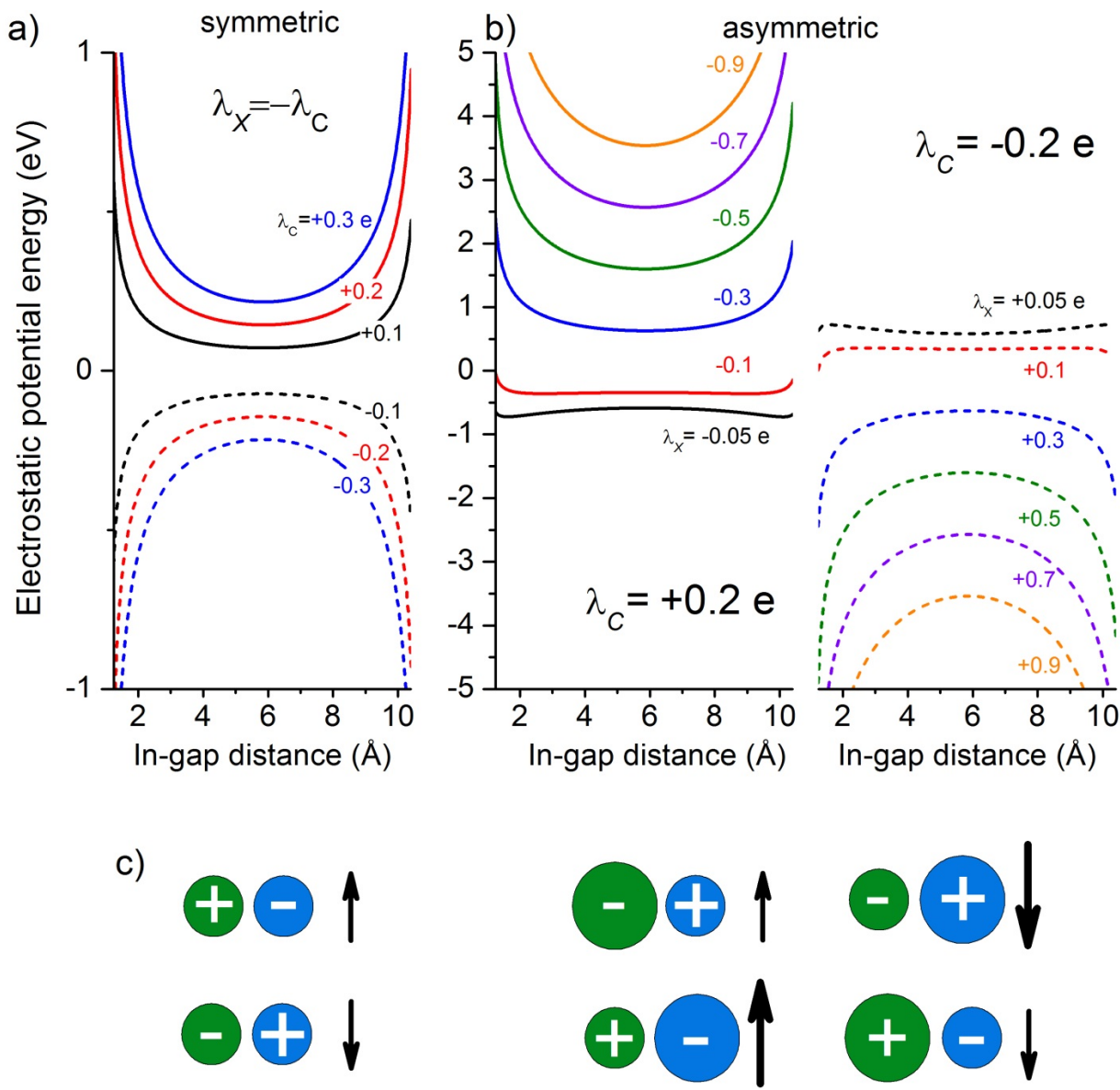

**Figure S7**. Electrostatic potential energy $E_p$ of four-line model (Figure 3b) for a) symmetric dipole model, i.e. $\lambda(X) = -\lambda(C)$, and b) asymmetric dipole model i.e. $\lambda(X) \neq -\lambda(C)$. In a) $\lambda(C)$ carries charge from -0.3 e to +0.3 e. In the left panel of b) $\lambda(C)$ is positive (+0.2 e) and $\lambda(X)$

varies from -0.05 e to -0.9 e, while in the right panel $\lambda(C)$ is negative (-0.2 e) and $\lambda(X)$ varies from +0.05 e to +0.9 e. c) In the symmetric case ($\lambda(X) = -\lambda(C)$, left panel), negative $\lambda(X)$ (green circles) raises in-gap potential energy (upward arrow), while positive $\lambda(X)$ decreases the energy (downward arrow). In the asymmetric case (right panel), the sign of total charge $\lambda(X) + \lambda(C)$ determines whether the $E_p$ is rising ($\lambda(X) + \lambda(C) < 0$) or lowering ($\lambda(X) + \lambda(C) > 0$). The effect is more pronounced (larger arrows) for larger $\lambda(X)$ (blue circles).

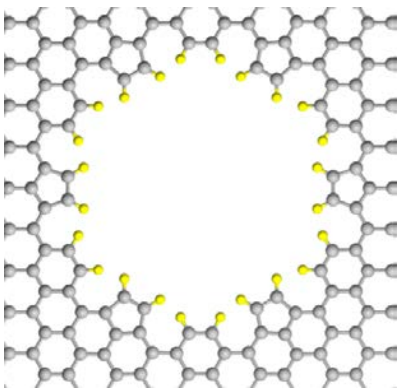

**Figure S8.** Geometry of H-terminated nanopore used in GPAW [1] calculations (H atoms are in yellow, while C atoms are grey).

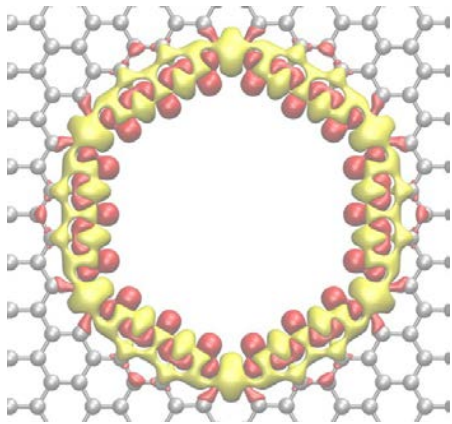

**Figure S9.** Isocontour plot of N-induced electron density upon the edge of graphene nanopore was passivated with nitrogen. Red and yellow isocontours correspond to the charge accumulation and depletion, respectively.